# Prediction of the Number of COVID-19 Confirmed Cases Based on K-Means-LSTM

Shashank Reddy Vadyala[1]*, Sai Nethra Betgeri[1], Eric A. Sherer[2], Amod Amritphale[3]

1. *Department of Computational Analysis and Modeling, Louisiana Tech University, Ruston, LA United States*
2. *Department of Chemical Engineering, Louisiana Tech University, Ruston, LA United States*
3. *Department of Internal Medicine, Division of Cardiology, University of South Alabama, Mobile, AL United States*

Correspondence should be addressed to Shashank Reddy Vadyala; srv009@latech.edu

**Abstract**

**Background:**

COVID-19 is a pandemic disease that began to rapidly spread in the US with the first case detected on January 19, 2020, in Washington State. March 9, 2020, and then increased rapidly with total cases of 25,739 as of April 20, 2020. The Covid-19 pandemic is so unnerving that it is difficult to understand how any person is affected by the virus. Although most people with coronavirus 81%, according to the U.S. Centers for Disease Control and Prevention (CDC), will have little to mild symptoms, others may rely on a ventilator to breathe or not at all. SEIR models have broad applicability in predicting the outcome of the population with a variety of diseases. However, many researchers use these models without validating the necessary hypotheses. Far too many researchers often "overfit" the data by using too many predictor variables and small sample sizes to create models. Models thus developed are unlikely to stand validity check on a separate group of population and regions. The researcher remains unaware that overfitting has occurred, without attempting such validation. In the paper, we present a combination algorithm that combines similar days features selection based on the region using Xgboost, K-Means, and long short-term memory (LSTM) neural networks to construct a prediction model (i.e., K-Means-LSTM) for short-term COVID-19 cases forecasting in Louisana state USA. The weighted k-means algorithm based on extreme gradient boosting is used to evaluate the similarity between the forecasts and past days. The results show that the method with K-Means-LSTM has a higher accuracy with an RMSE of 601.20 where as the SEIR model with an RMSE of 3615.83.

## Introduction

In December 2019, unknown viral pneumonia was identified in the City of Wuhan, China [1]. In February 2020, the WHO identified it as a coronavirus – and named it COVID-19 which presents as an acute respiratory syndrome and a highly communicable disease that is extremely efficient at the person to person transmission. According to guidance for diagnosis and treatments for COVID-19 issued by the World Health Organization, COVID-19 was transmitted through respiratory aspirates, droplets, contacts, and feces, and aerosols transmission is highly possible[2]. Since its initial detection, the COVID-19 has spread rapidly; it has been declared as a pandemic and has infected over 5.2 million people in more than 210 countries [3-4]. Despite measures implemented in China during the initial phase of spread, multiple epicenters have emerged across the world especially in certain European countries and in the United States of America (USA). The USA has 1.6 million confirmed cases and 96,662 deaths, the most of any country in the world, as of the date _according to the World coronavirus update [4]. In the USA, the State of Louisiana has 25,739 confirmed cases (9[th] highest state) and 1,540 deaths (8[th] highest state) as of the date [5]. The mean age of these who died in Louisiana was 70 years, and lots of patients had diabetes 36.65% and cardiovascular diseases 20.92% [6]. On March 23, 2020, the Governor of Louisiana implemented extraordinary measures to limit the transmission by implementing the Stay at Home order that intended to scale back the people are available in contact with those that are infected. This decision was made in light of a daily rate of infection consistently between 1.5% and 3.5% with an overall average of 2.6% from March 9th until May 13, 2020, and the proportion of patients admitted to intensive care units consistently between 14 to 16% who were actively infected [7].

The rapid spread of COVID-19 has highlighted the need to understand how population dynamics interact with pandemics [8]. For example, the effects of the virus are more severe in the elderly and population aging is currently more pronounced in wealthier countries which potentially may lessen the impact of this pandemic in lower-income countries with weaker health systems, but younger age structures. Climate variables can also be a direct cause of biological interactions between COVID-19 and humans. Optimal temperature, humidity, and wind speed are variables that can determine the survival and transmission of the SARS virus [9-11]. Changes in weather are very significantly correlated with changes in mortality rates due to pneumonia [12]. The extreme weather conditions that accompany long-term climate change may also contribute to the spread of the West Nile virus in the United States and Europe[13]. Certain climatic conditions can be considered as the top predictors of respiratory diseases such as SARS.

Several statistical and mathematical models, such as the SEIR [14] (susceptible, exposed, infected, and resistant) model, have been designed to simulate the effect of disease on many levels. In the SEIR model, the transmission of disease or the incubation period is incorporated. The SEIR model is useful for estimating the dynamics of transmission but the model rests on simplifications. These models are usually written as ordinary differential equations (ODE) as opposed to more comprehensive models of the nonlinear dynamics comprehensively that include the time-variant and probabilistic variables into the models. It is necessary to define all the parameters for estimation like rate of transmission of disease. The data scarcity renders at an early stage models that are unfeasible to predict the trend disseminate. The accuracy of forecasting the future cases of COVID-19 can be improved relative to the SEIR model can be improved by including these features in a model.

To overcome limitations of the SEIR model approach, and assist public health planning and policymaking, we are using on deep learning models for real-time forecasting of the new and cumulative confirmed cases of COVID-19 in total across Louisiana state. Viruses can be transmitted by being influenced by several factors, including climatic conditions (such as temperature and humidity), and population density, race but few models include the effects of climate, population density on COVID-19. In this study, we compare the performance of K-Means-LSTM models to more traditional compartmental models (SEIR) for estimating the spread of an infectious disease outbreak. The deep learning model is equipped with individual parish, demographics, population density, age, temperature, humidity, average household income per year, and risk of COVID-19 for the parish to understand these effects on COVID-19 transmission and to produce more accurate predictions.

The structure of this paper is as follows: Section 2 introduces the data source. Section 3 gives a brief description of the epidemiologic transmission model used for forecasting. Section 4 introduces the proposed method—section 5 comparative analysis of different experimental results. Section 6 is the Discussion and conclusion.

**Methods**

**Datasources**

Data on the COVID-19 pandemic is available from the GitHub repository managed by the John Hopkins University was collected for the modeling. This is a collection of publicly available data from multiple sources, which are processed and delivered by the Johns Hopkins University Center for Systems Science and Engineering (JHU CSSE). The COVID-19 Data (by John Hopkins University) contains the daily number of confirmed deaths, and recovered people. The Basic information contains information about the date, country (Reconfirm: country or county), and state of the cases. Notice that the data are provided to the public strictly for educational and academic research purposes. In this study, we are focus on the state of Louisiana State daily figures and the three variables of deaths, confirmed cases, and recoveries. The data refer to daily cumulative cases and cover the period from January 22, 2020, until April 19, 2020. The data are updated daily, and the files used in this paper have been downloaded from [15]

Demographic data for the state of Louisiana was downloaded [16]. The following variables were collected: parish name, parish population, parish population density, race (white, African American, Asian, other, or unknown), ethnicity (Hispanic or non-Hispanic), age by categories, average daily temperature, average daily humidity, and the median age in the parish. In the analysis, we included only parish demographics data, which were reported with COVID-19 cases repository managed by John Hopkins University.

The weather data is included in this research for the period of research were obtained from the [17]The data consist of temperature minimum (°F), temperature maximum (°F), temperature average (°F), and humidity (%).

**Data Preparation.**

**Our analysis is based on the following information:**

- The daily reported the number of confirmed cases from 3rd March 2020 to 26th May 2020 in Louisiana state parishes.
- Demographic information of these Louisiana state parishes.
- Temperature and humidity during our study period at these Louisiana state parishes.
- Combined all three data into one final dataset.

**Data Analysis**

Pearson correlation test was used to examine the relationship between weather and daily COVID-19. Table 1 shows that, among four weather variables, only temperature average (°F) significantly correlated with COVID-19 ($r = 0.439; p < .001$), with a medium level. Temperature minimum, temperature maximum, and humidity were not significantly correlated with COVID-19. So we used only Temperature avg °F from an all-weather variable for further analysis.

**Table 4.** Pearson correlation coefficients between COVID-19 and weather variables

| Weather variable | Pearson correlation coefficient |
|---|---|
| Temperature min °F | 0.189 |
| Temperature max °F | 0.245 |
| Temperature avg °F | 0.439 |
| Humidity (%) | 0.004 |

**SEIR Model**

Kermack and McKendrick (1927) first proposed the SEIR type epidemiological models for the modeling of infectious disease studies. SEIR models people are divided into population four states: susceptible (S), exposed/latent (E), infectious (I), and recovered/removed (R) individuals and models are applied to based on the classification of the contamination of individuals involved.

The population size is N where N = S + E + I + R. Here we normalize to one, so all results should be interpreted infractions of the relevant state S = S/N, I = I/N, E=E/N, and R = R/N. The dynamics between the four states are described by differential equations (4) – (6), and the transfer rates between these four states decide how an epidemic plays out over time. Where σ is the incubation rate (the latent rate at which a person becomes infectious); γ is the recovery rate (determined by 1/D where D is the duration of infection); ξ the rate at which recovered people become susceptible (due to low immunity or other health-related issues), and β is the infectious rate (the probability of disease-transmitting to susceptible persons from an infectious person).

$$\frac{ds(t)}{dt} = -\beta I(t) S(T) \tag{1}$$

$$\frac{dE(t)}{dt} = \beta I(t) S(T) - \xi E(T) \tag{2}$$

$$\frac{dI(t)}{dt} = \xi E(t) - \Upsilon I(t) \tag{3}$$

$$\frac{dR(t)}{dt} = \Upsilon I(t) \tag{4}$$

In this model, the $R_0$ value is important. It tells about the contagiousness of the disease. $R_0$ determines an average of what number of people can be affected by a single infected person over time. $R_0$ It depends on the exponential rate of growth of the COVID−19 outbreak, as well as additional variables such as the latent duration (time from infection) and infectious duration, all of which can not be determined directly from the results. $R_0$ can be calculated using the ratio of the transmission rate to the recovery rate shown in equation (5).

$$R_0 = \frac{\beta}{\Upsilon} \tag{5}$$

**Table 1:** Modeling Assumptions and ParameterSetting:

| Symbol | Description | Estimate Mean Value | Data Source |
|---|---|---|---|
| S | Susceptible individuals | - | |
| E | Exposed individuals in the latent period | - | |
| I | Infectious individuals | - | |
| R | Recovered individuals with immunity | 0.05 | [18, 19] |
| $\beta$ | Infection rate | 0.040 | [18, 19] |
| $1/\Upsilon$ | Average infectious period | 3.85 | [18, 19] |
| N | Total population size | Average Population of the parish | [16] |
| $R_0$ | Basic reproduction number | 4.83 | [18, 19] |
| $1/\xi$ | Average latent period | 1/5 | [18, 19] |

**The proposed framework**

In this section, we first present the overview of our proposed deep learning architecture and then discuss each component of the forecasting framework process is shown in Figure. 1.

**Figure 1**. Overall Procedure of Forecasting COVID-19 Cases.

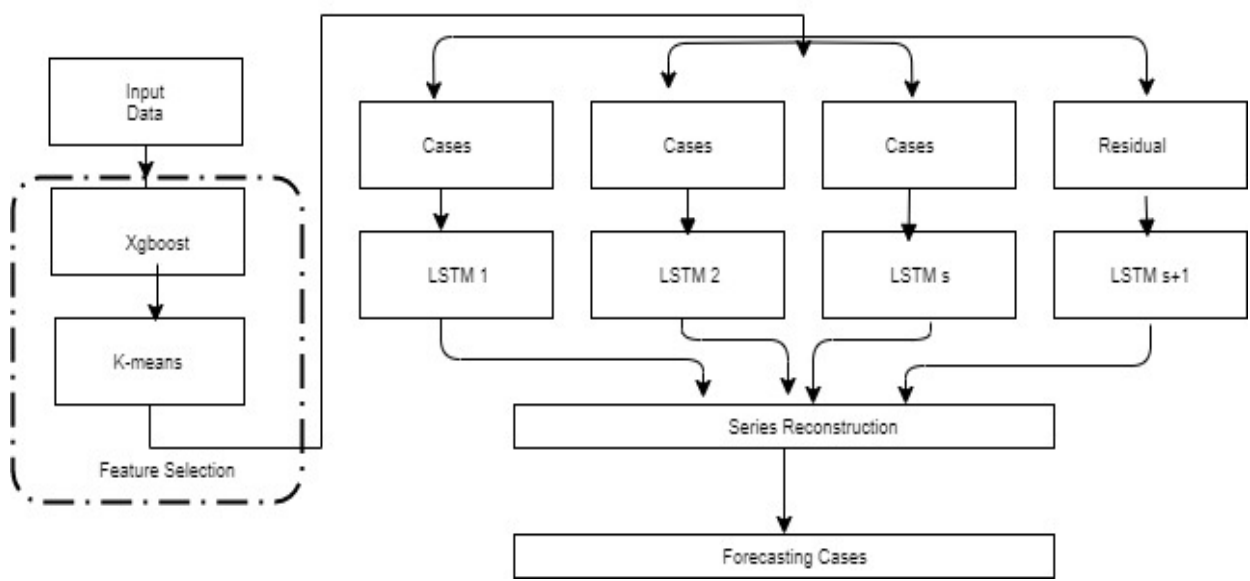

**The Overall Procedure**

First, we estimate the probability of COVID-19 from age and race for each parish in Louisiana mentioned below. Then, a clustering algorithm is applied to the features to find the optimal grouping of time series data of COVID-19 confirmed cases and parishes. After clustering, each cluster of time series data of COVID-19 confirmed cases and parishes would go through a series of preprocessing steps like normalization, as shown in equation (6) to generate input data for the LSTM model. In the following sections, we explain a brief introduction to the LSTM model and time series clustering using K-means algorithm.

$$X_{scaled} = \frac{X - X_{min}}{X_{max} - X_{min}} \tag{6}$$

**Table 2.** Features used for predicting new COVID-19 cases.

| Features | Description | Symbol |
|---|---|---|
| Date | Date | $D_t$ |
| Parish | Parish Name | $P$ |
| Population Density | Population Density in the parish | $P_d$ |
| Race | Total number of individuals of the race k | $R_k$ |
| Median Age | The median age in the parish | $A_{avg}$ |
| Temperature min °F | Minimum Temperature in the parish | $T_{min}$ |
| Temperature max °F | Maximum Temperature in the parish | $T_{max}$ |
| Temperature avg °F | Average Temperature in the parish | $T_{max}$ |
| Humidity (%) | Humidity in parish | $H$ |
| Median Household income per year | Median Household income per year in the parish | $I_{avg}$ |
| Confirmed COVID-19 cases | COVID-19 Incidence of prediction day in the parish | $C$ |
| Risk Of COVID-19 | Probability of risk of COVID-19 in the parish | $P_c(A_{avg}, R_k)$ |

**The complete procedure of the proposed K-Means LSTM model is presented as follows:**

- We combined days selection. Calculate the weight of the feature by xgboost method, then combined with the K-means algorithm to determine the similar day's cluster.
- Separated LSTM neural networks employed to forecast COVID-19 cases for each day and parish, respectively. Reconstruction of the predictive values for each single LSTM model.

Estimating the Probability of COVID-19 from age and race

COVID-19 case rates are available either by age or by race, but not for each and race mix. We model the model rate with a logistic to obtain these estimates between age brackets(e.g., 10-20, 20-30,… etc.) and race (white, African American, Asian, other, or unknown), ethnicity (Hispanic or non-Hispanic). This model gives the probability of risk of COVID-19 $P_c(A_{avg}, R_k)$ for each parish as shown in equation (7).

$$P_c(A_{avg}, R_k) = \sigma(\delta_{age}(a_i) + \delta_{race}[race \,\epsilon\, r_i]) \tag{7}$$

Current data sources say that probabilities $P_c(A_{avg})$ and $P_c(R_k)$ are independent [20]. To assume a mutual distribution, we assume a linear (logistic) relationship between age group, different races: Where Where $\delta_{age}(a_i)$ has a value for each age group and $\delta_{race}$ are scalars. The marginal distributions $P_c(A_{avg})$ and $P_c(R_k)$ are reported by Louisiana Facility [21]. We assume that these distributions are the same in COVID-19 patients as in the general population. [21]

**Feature Selection:**

To improve learning performance and quality extraction features, data must be preprocessed and more features displayed to promote model learning. Feature extraction is used to compress large data sets utilizing dimensionality reduction procedure in time series data mining. Using feature extraction, computational efficiency can be increased and the use of more sophisticated algorithms is possible. Most methods of extraction of the features are standardized in nature; the features extracted are typically based on application. Therefore, one set of features that work well on one application may not be relevant to another application. The identification feature involves the application of Datamining techniques like clustering algorithms to derive usage patterns from the data. Extracting the simplified characteristics of data from time series will offer a more significant reduction in dimensionality (Features) compared to other approaches.

This section presents an alternative to conventional feature selection by calculating the weights of the feature using the Xgboost algorithm and combining the weighted feature using the clustering of the k-means algorithm. Three ensemble implementations in common packages such as XGBoost [22], scikitlearn [23], and the python package [24] enable a user to use clustering and XGBoost to compute a measure of feature importance.

**Feature-Weight Learning Algorithm: Extreme Gradient Boosting**

Xgboost [25] is an improved decision-making algorithm based on the gradient boosting tree and can create boosted trees efficiently and run in parallel.The core of the algorithm is to optimize the value of the objective function. Unlike the use of the feature, vectors to calculate the similarity between the forecasting and history days, gradient boosting constructs the boosted trees to intelligently obtain the feature scores, thereby indicating the importance of each feature to the training model.

The value of a feature depends on how the predictive output changes dramatically when repla cing such a feature with random noise. With the COVID-19 instances, we take multiple features as input for the Xgboost algorithm to measure the functional significance. In the Xgboost algorithm's training course, we can obtain how each feature contributes to the prediction results. COVID-19 cases are sensitive to Risk of COVID-19 of the parish for COVID-19 case forecasting(shown in Figure 2). The essential values of all features that will be used as a priori knowledge of the subsequent clustering algorithm are now extracted from them.

**Figure 2.** XGBoost feature importance.

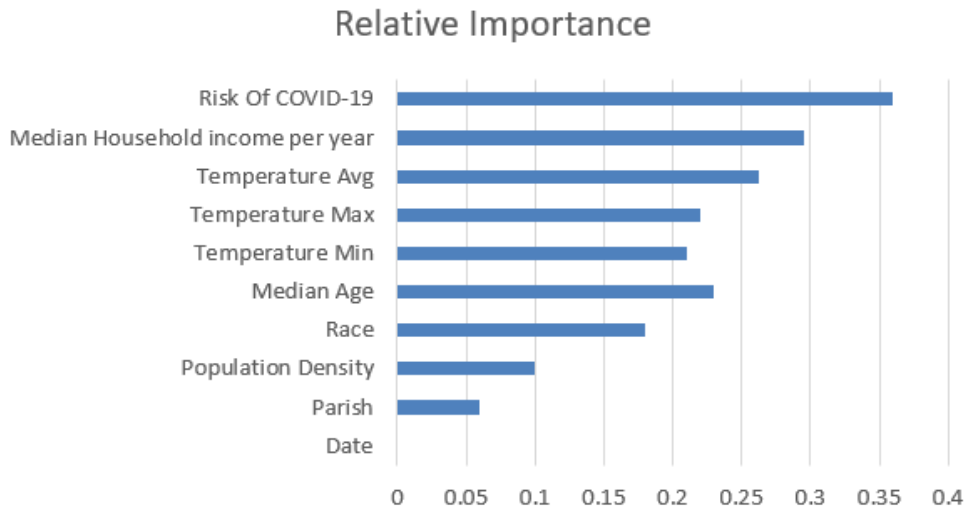

**K-means:**

Clustering are the most useful tools for data mining, compression, probability density estimation, and many other essential tasks. There are many clustering algorithms like K-means, PAM, DBSCAN, many more. K-means algorithm is the most used clustering method. J. Macqueen originally proposed the K-Means algorithm in 1967 [26].

**The steps of using clustering for feature selection are presented as follows.**

- The input data is normalized
- The forecasting day is selected as the first center $D_0$.
- The next center $u_j$ is selected, where $u_j$ Is the most distant point from the cluster centers previously identified { $u_0, u_1, u_2,,,,, u_{t-1}$ }. Steps 2 and 3 will be repeated until the centers K are identified.
- The weights of the features are calculated by using the algorithm Xgboost. After that, the weights are attributed to each feature, thereby providing them with different levels of importance. Let $w_p$ be the weight is associated with the feature p. The norm is presented as follows.

$$f(d_i, d_j) = \sqrt{w_1(d_{i1} - d_{j1})^2 + \ldots + w_n(d_{in} - d_{jn})^2} \qquad (8)$$

  a. That data point is allocated to the closest cluster.
  b. Updating the clusters is achieved by recalculating the centroid cluster. The algorithm executes (a) and (b) repeatedly until convergence is attained.

Our proposed method Xgboost-k-means can more easily combine features into one cluster than the simple algorithm k-means does. The features may then be the input data for subsequent loading to forecast the number of cumulative COVID cases. For forecasting the number of confirmed, features were extracted from raw data, as mentioned in the data source. We extracted 12 features from the data. Deaths, confirmed cases, recovered. Parish name, the population of the parish, population density of the parish, race (aggregated into white, African

American, Asian, other, and unknown), ethnicity (Hispanic or non-Hispanic) and median age. Temperature average (°F).

**LSTM Model:**

The LSTM is an improvement on RNN [27], it contains a processor that determines whether the information is useful or not, in which the working part is named as a cell. There are three doors in a cell: the input layer door, the forget-gate, and the output layer door. Input layer door and forget-gate both work on the state of cells. However, the role of the input gate is to selectively record new information into the cell state, while forget-gate is aimed at selectively forgetting information about cell states. The output layer gate acts on the hidden layer to output information. The structural model is shown in Figure. 3. In this study, the aim was to estimate the number of positive COVID-19 cases through time; as this is a well-suited task for the LSTM model, we used this model in our study.

Figure 3. LSTM cell diagram.

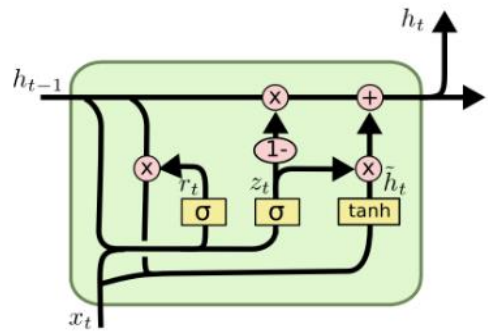

Figure 4. LSTM neural networks model for daily COVID-19 Case forecasting.

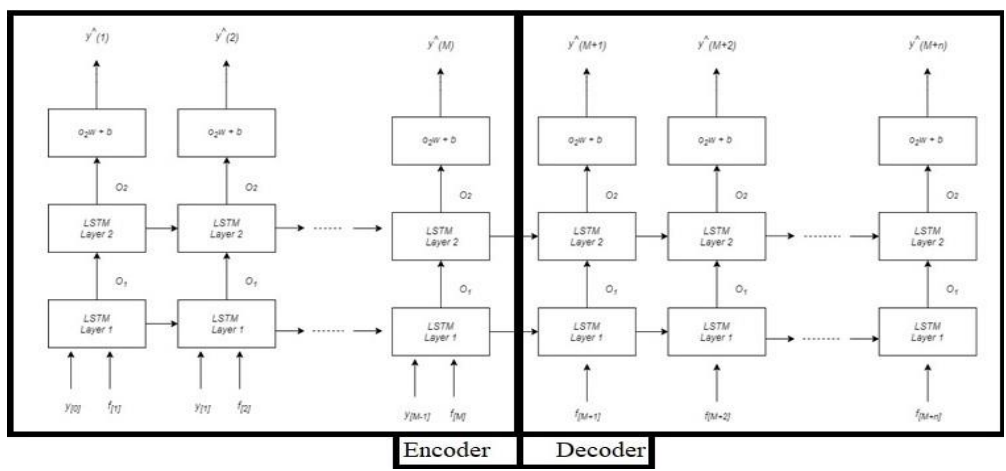

The architecture is divided into two parts: 1) Encoder 2) Decoder shown in Figure 4. The encoder aims to prediction the COVID-19 cases we have already known, yet $y_{[M-1]}$ and

$f_{[M]}$ present the COVID-19 cases of the previous day and the features of this day, respectively. The decoder generates an output sequence $\hat{y} = \{\hat{y}_{|M|}, \hat{y}_{|M+1|},..,\hat{y}_{|M+n|}\}$ that is the forecast of the COVID-19 cases for the next n+1 day. All the separate LSTM for each parish forecasts are added to get total number of COVID-19 cases.

**Parameter Setting and Training**

Each city in Louisiana was trained using its case data. Because of the relatively small dataset, we developed a more straightforward network structure to prevent overfitting. The model was optimized using Adam optimizer and ran for 500 iterations. Although during optimization, the network was validated by predicting active cases, the same network was used to predict death, recovery, and the current number of cases. The COVID-19 cases from March 9, 2020, until May 19, 2020, were used for tests, and rest were used for training. Standard backpropagation is used to train the network using Stochastic Gradient Descent(SGD), a gradient-based technique. We used the Keras [28] library with the TensorFlow backen [29] to train the proposed models.

**Table 3.** Parameters used in optimum.

| Parameter | Selection | Selection |
|---|---|---|
| Learning rate | Log uniform | 1e-1 to 1e-7 |
| Hidden layers | Discrete numeric | 1 to 20 |
| Hidden state | Discrete numeric | 1 to 200 |
| Activation | Category | {ReLu,sigmoid, tanh} |
| Batch size | Discrete numeric | 1 to 10 |
| Dropout | Log uniform | 0 to 0.5 |

**Result**

We compared our K-means-LSTM prediction results with the results of the SEIR model are shown in Figure 6. To accurately assess the performance of the SEIR model, K-means-LSTM and LSTM, Root means square error(RMSE) were calculated and shown in Table 5, and their formulae presented in Equation(9) using pyton package[30]

$$\boldsymbol{RMSE} = \sqrt{\sum \frac{(y_{pred} - y_{actual})^2}{N}} \quad (9)$$

**Confirmed Cases:** First COVID-19 case was reported in Louisiana on March 9[th], 2020. Reports the growth rate of cases on that specific date shown in Figure 5. There are several dates with outlier growth rates due to testing rollouts. Louisiana is leading in the per capita test score, relative to the rest of th United States. 75,000 individuals have been studied in a population of nearly 4.7 million people. . Every day, Louisiana received around 5,600 new tests, with 6.8 percent of those showing positive results. A total of 306 new cases reported by the State per day. [31]

**Figure 5.** The cumulative number confirmed cases of COVID-19 across Lousiana

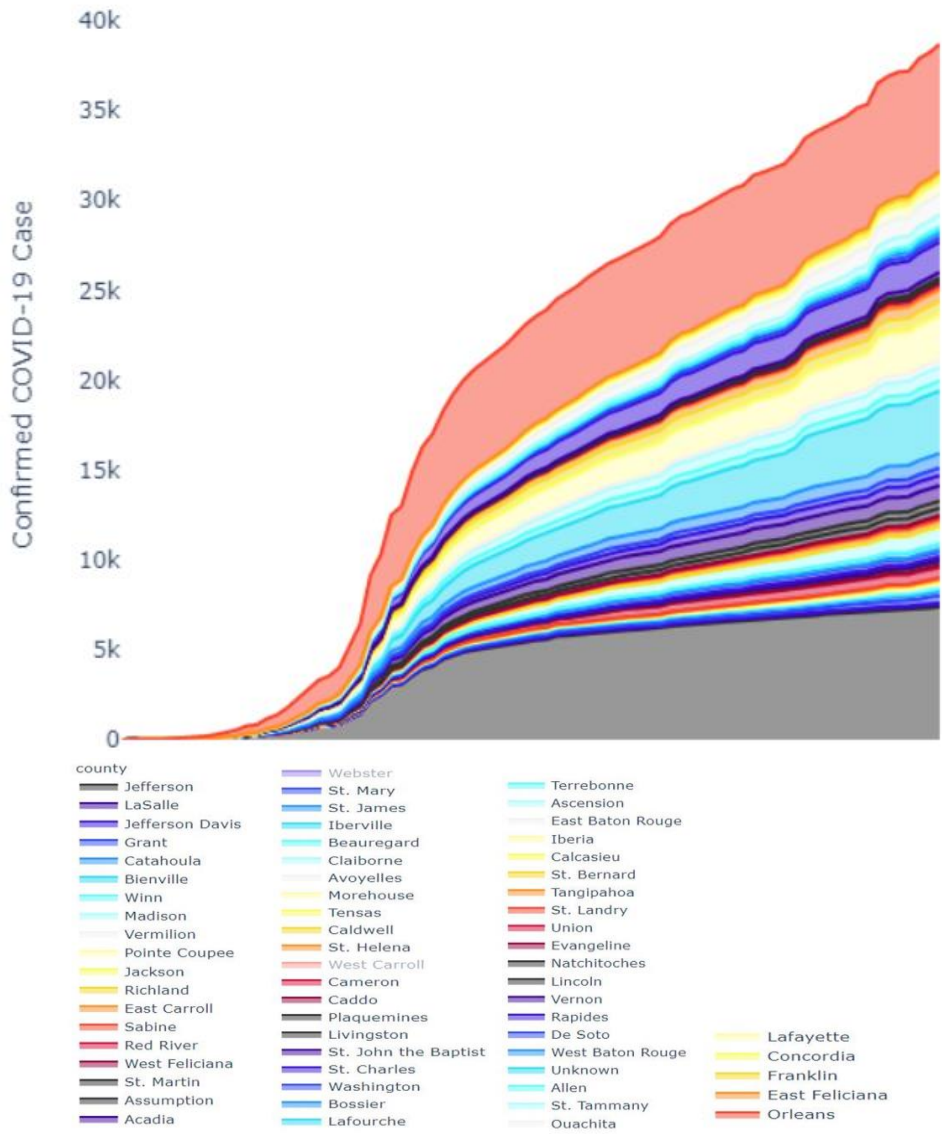

**Figure 6.** Comparisons between cumulative confirmed cases OF COVID-19 K-Means-LSTM COVID-19: coronavirus disease.

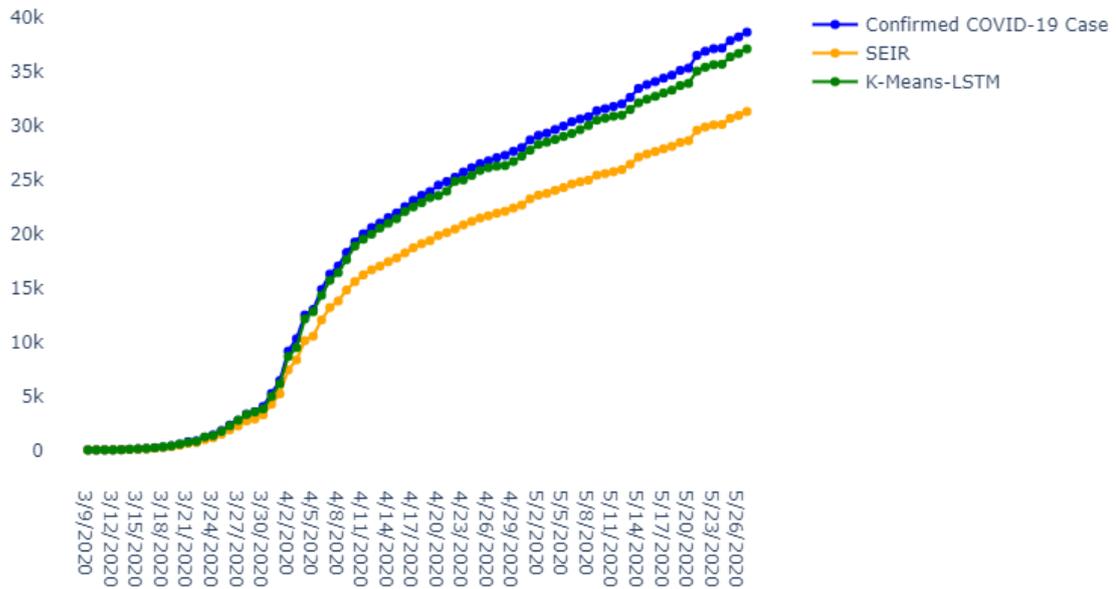

The K-Means-LSTM model was evaluated for different parishes. The proposed K-Means-LSTM can fit the case count by parish well using demographics, population density, average age, temperature, humidity, average household income per year, and risk of COVID-19 for the parish. In general, the model fits well were the case counts(data points) are higher than 45 and as data becomes richer, the fit improves significantly.

**Figure 7.** Shows cumulative data points for each parish in Lousiana

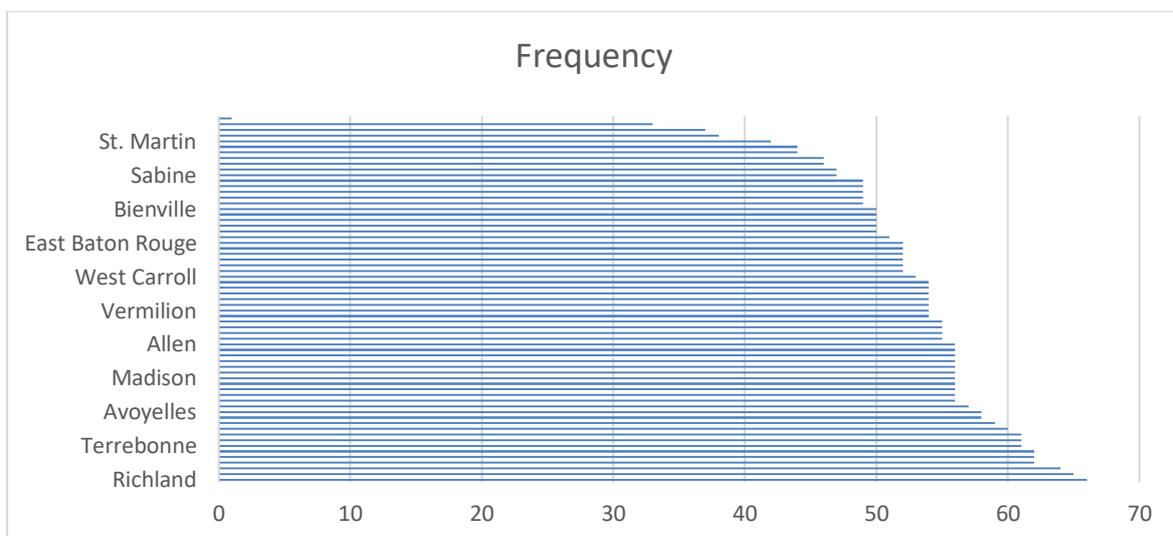

The comparison has also been made for the confirmed COVID-19 cases, predicted COVID-19 cases with K-Means-LSTM and SEIR from 5th to 13th May 2020. Out of 64 parishes, three parishes are selected based on data points are shown in Table 4(a)- 4(c)
.

| Table 4(a): Forecast of the number of confirmed cases in Richland, Louisiana. (Data points=65) | | | |
|---|---|---|---|
| Date | Real | K-Means-LSTM | SEIR |
| 5/7/2020 | 91 | 93 | 69 |
| 5/8/2020 | 95 | 94 | 75 |
| 5/9/2020 | 99 | 96 | 79 |
| 5/10/2020 | 100 | 102 | 83 |
| 5/11/2020 | 103 | 103 | 86 |
| 5/12/2020 | 103 | 106 | 91 |
| 5/13/2020 | 104 | 110 | 93 |

| Table 4(b): Forecast of the number of confirmed cases in St.Martin, Louisiana.(Data points= 39) | | | |
|---|---|---|---|
| Date | Real | K-Means-LSTM | SEIR |
| 5/7/2020 | 254 | 209 | 129 |
| 5/8/2020 | 255 | 218 | 136 |
| 5/9/2020 | 257 | 221 | 149 |
| 5/10/2020 | 257 | 229 | 152 |
| 5/11/2020 | 260 | 232 | 163 |
| 5/12/2020 | 264 | 239 | 175 |
| 5/13/2020 | 276 | 241 | 192 |

| Table 4(c): Forecast of the number of confirmed cases in Calcasieu, Louisiana.(Data points = 23) | | | |
|---|---|---|---|
| Date | Real | K-Means-LSTM | SEIR |
| 5/7/2020 | 478 | 402 | 326 |
| 5/8/2020 | 481 | 411 | 331 |
| 5/9/2020 | 498 | 422 | 341 |
| 5/10/2020 | 501 | 437 | 357 |
| 5/11/2020 | 508 | 461 | 369 |
| 5/12/2020 | 512 | 472 | 375 |
| 5/13/2020 | 537 | 506 | 381 |

Table 5: Experimental results in terms of RMSE

| Parish | K-Means-LSTM | SEIR |
|---|---|---|
| Louisiana | 601.20 | 3615.83 |

**Discussion**

It takes effective strategies to prevent and control the epidemic spread. Estimating the epidemiological pattern of outbreak prevalence is critical for the allocation of medical services, for controlling manufacturing activities, and even for the country's national economic growth. Thus, it is essential to create a reliable and suitable forecasting model that can help governments as a reference to decide on emergency macroeconomic strategies and medical resource allocation. SEIR model is one of the most commonly used epidemiological models.

Although SEIR-based models were widely used to model the COVID-19 outbreak, they do contain some degree of uncertainty. As an alternative to the SEIR model, this research proposed a deep learning model (K-Means-LSTM) as a new trend in advancing outbreak modeling. The K-Means-LSTM strategy does not predict the infection's pandemic and spread. Instead, it predicts the infected cases time series. The K-Means-LSTM is used for this analysis to forecast the COVID-19 outbreak for Lousiana. This study suggests K-Means-LSTM as a useful tool for modeling the outbreak based on the results reported here and given the complex nature of the COVID-19 outbreak and its parish-to-parish behavioral variation. Training data is used to train the algorithm and ascertain the best set of parameters to be used in K-Means-LSTM. K-Means-LSTM has three main advantages. First, the model learns not only from the historical COVID-19 data but also from COVID-19 's applied demographic data, weather and risk for each parish. The model's second advantage is its capacity to be applied at different scales. It can currently forecast the spread in a parish and state. Lastly, the model can forecast short and long-term forecasts that could be a reliable decision-making tool.

**Limitation**

In comparison, we note that the width of prediction intervals made by K-Means-LSTM decreases on average as more data is used for predictions. It can partially be attributed to the larger number of cases and the longer initial forecast time seen in many parishes. K-Means-LSTM outperforms the commonly used SEIR, but more features and input data are needed to estimate the trends. Although we have promising results for data from Louisiana, these new approaches need to be further assessed on other datasets. However, the findings presented are encouraging and will inspire the group to adopt these new tools quickly.

**Conclusion**

Accurate COVID-19 case forecasting is a significant problem for public health authorities to efficiently and timely coordinate patient care and other services required to solve the epidemic. In this research, we propose a K-Means-LSTM neural network to tackle the issue of variance and precision in predicting the number of reported cases in the traditional SEIR model. The findings of the study will help policy and healthcare efficiently prepare and provide services to handle the situation in these states over the next few days and weeks, including nurses, beds, and intensive care facilities. The data should be updated in real-time for more precise comparison and future perspectives.

**Abbreviations**

- COVID-19: coronavirus disease
- LSTM: long short-term memory
- RMSE: root mean square error
- SEIR: Susceptible Exposed to Infectious Recovered
- WHO: World Health Organization
- USA: United States of America